\documentclass[12pt]{article} \hoffset=-15mm \voffset=-10mm
\usepackage{graphicx,amsmath,amssymb,epsf} 
\usepackage{latexsym,bm,slashed} 
\usepackage{xcolor} \definecolor{dark}{rgb}{0.10,0.2,0.3}
\definecolor{magenta}{rgb}{0.7,0.1,0.3}
\definecolor{purpure}{rgb}{0.5,0.15,0.3}
\usepackage[font=small,format=plain,labelfont=bf,up,textfont=it,up]{caption}
\usepackage{hyperref, cite} \hypersetup{colorlinks, citecolor=blue,
  filecolor=blue, linkcolor=magenta,
  urlcolor=purpure,hyperfootnotes=true,pdftex}

\def\be{\begin{equation}}
\def\ee{\end{equation}}
\def\bea{\begin{eqnarray}}
\def\eea{\end{eqnarray}}

\def\ubar{\bar{u}}

\newcommand{\slv}{\raise.15ex\hbox{$/$}\kern-.53em\hbox{$v$}}
\newcommand{\sln}{\raise.15ex\hbox{$/$}\kern-.53em\hbox{$n$}}
\newcommand{\slnbar}{\raise.15ex\hbox{$/$}\kern-.53em\hbox{$\bar{n}$}}
\newcommand{\slF}{\raise.15ex\hbox{$/$}\kern-.53em\hbox{$F$}}
\newcommand{\sll}{\raise.15ex\hbox{$/$}\kern-.40em\hbox{$l$}}
\newcommand{\slh}{\raise.15ex\hbox{$/$}\kern-.40em\hbox{$h$}}
\newcommand{\slP}{\raise.15ex\hbox{$/$}\kern-.53em\hbox{$P$}}
\newcommand{\slp}{\raise.15ex\hbox{$/$}\kern-.53em\hbox{$p$}}
\newcommand{\slq}{\raise.15ex\hbox{$/$}\kern-.53em\hbox{$q$}}
\newcommand{\slR}{\raise.15ex\hbox{$/$}\kern-.53em\hbox{$R$}}
\newcommand{\slz}{\raise.15ex\hbox{$/$}\kern-.53em\hbox{$Z$}}
\newcommand{\slzbar}{\raise.15ex\hbox{$/$}\kern-.53em\hbox{$\bar{Z}$}}
\newcommand{\slQ}{\raise.15ex\hbox{$/$}\kern-.53em\hbox{$Q$}}
\newcommand{\slK}{\raise.15ex\hbox{$/$}\kern-.53em\hbox{$K$}}
\newcommand{\slk}{\raise.15ex\hbox{$/$}\kern-.53em\hbox{$k$}}
\newcommand{\slkbar}{\raise.15ex\hbox{$/$}\kern-.53em\hbox{$\bar{k}$}}
\newcommand{\slkone}{\raise.15ex\hbox{$/$}\kern-.53em\hbox{$k_1$}}
\newcommand{\slpone}{\raise.15ex\hbox{$/$}\kern-.53em\hbox{$p_1$}}
\newcommand{\slpbarone}{\raise.15ex\hbox{$/$}\kern-.53em\hbox{$\bar{p}_1$}}
\newcommand{\slptwo}{\raise.15ex\hbox{$/$}\kern-.53em\hbox{$p_2$}}
\newcommand{\slpbartwo}{\raise.15ex\hbox{$/$}\kern-.53em\hbox{$\bar{p}_2$}}
\newcommand{\slqone}{\raise.15ex\hbox{$/$}\kern-.53em\hbox{$q_1$}}
\newcommand{\slD}{\raise.15ex\hbox{$/$}\kern-.53em\hbox{$\!D$}}
\newcommand{\slC}{\raise.15ex\hbox{$/$}\kern-.53em\hbox{$C$}}
\newcommand{\slA}{\raise.15ex\hbox{$/$}\kern-.73em\hbox{$A$}}
\newcommand{\slSigma}{\raise.15ex\hbox{$/$}\kern-.53em\hbox{$\Sigma$}}
\newcommand{\slpartial}{\raise.15ex\hbox{$/$}\kern-.53em\hbox{$\partial$}}
\newcommand{\slcalP}{\raise.15ex\hbox{$/$}\kern-.63em\hbox{$\cal P$}}
\newcommand{\sleps}{\raise.15ex\hbox{$/$}\kern-.53em\hbox{$\epsilon$}}
\newcommand{\slepsbar}{\raise.15ex\hbox{$/$}\kern-.53em\hbox{$\overline{\epsilon}$}}
\newcommand{\slepsstar}{\raise.15ex\hbox{$/$}\kern-.53em\hbox{$\epsilon$}^\star}
\newcommand{\slS}{\raise.15ex\hbox{$/$}\kern-.73em\hbox{$S$}}

\newcommand{\nn}{\nonumber\\}

 \title{\bf
  \Large \bf \Large Quark jets scattering from a gluon field: from saturation to high $p_t$} \author{ Jamal~Jalilian-Marian$^{1,2}$
  \bigskip \\
  {\normalsize $^1$Department of Natural Sciences, Baruch College,
    CUNY,}
  \\
  {\normalsize 17 Lexington Avenue, New York, NY 10010, USA}\\
  {\normalsize $^2$CUNY Graduate Center, 365 Fifth Avenue, New York, NY 10016, USA}\\
   }

\begin{document}

\maketitle
\begin{abstract}
We continue our studies of possible generalization of the Color Glass Condensate (CGC) effective theory of high energy QCD to include the high $p_t$ (or equivalently large $x$) QCD dynamics as proposed in~\cite{jjm-elastic}. Here we consider scattering of a quark from both the small and large $x$ gluon degrees of freedom in a proton or nucleus target and derive the full scattering amplitude by including the interactions between the small and large $x$ gluons of the target. 
We thus generalize the standard eikonal approximation for parton scattering which can now be deflected by a large angle 
(and therefore have large $p_t$) and also lose a significant fraction of its longitudinal momentum (unlike the eikonal approximation). The corresponding production cross section can thus serve as the starting point toward derivation
of a general evolution equation that would contain DGLAP evolution equation at large $Q^2$ and the JIMWLK evolution equation at small $x$. 
This amplitude can also be used to construct the quark Feynman propagator which is the first ingredient needed to generalize the Color Glass Condensate (CGC) effective theory of high energy QCD to include the high $p_t$ dynamics. We outline how it can be used to compute observables in the large $x$ (high $p_t$) kinematic region where the  standard Color Glass Condensate formalism breaks down.  
\end{abstract}

\section{Introduction}
Twist expansion and collinear factorization approach~\cite{cteq} to particle production in QCD is a powerful 
and extremely useful formalism for particle production in high energy hadronic/nuclear 
collisions at high $p_t$. However it is not expected to be valid at high energy and/or for large nuclei where 
twist expansion breaks down due to high gluon density effects (gluon saturation). The Color Glass 
Condensate (CGC) formalism (see \cite{cgc-reviews} for reviews) is an effective field theory 
approach to QCD at high energies which relies on the fact that at high energy (or at small $x$) 
a hadron or nucleus wave function contains many gluons, referred to as gluon saturation~\cite{glr,mq}, 
and hence is a dense many-body system which is most efficiently described via semi-classical methods~\cite{mv}. 
The most significant aspect of CGC is perhaps the emergence of a dynamical scale, called the saturation 
scale $Q_s$ , which grows with energy (or $1/x$) and hence can be semi-hard. The CGC formalism thus 
can be used to compute quantities such as gluon multiplicities,... which are not amenable to the standard 
perturbative methods. Even though applications of the CGC formalism to hadronic/nuclear processes 
in a limited range of kinematics at RHIC and the LHC have been quite successful~\cite{lappi}, the CGC 
formalism has its shortcomings, namely it is not valid when one probes large $x$ modes of the target 
proton or nucleus. In case of particle production in hadronic collisions this happens when high $p_t$ 
particles are produced since $x$ and $p_t$ are kinematically related, $x \sim {p_t \over \sqrt{s}}$. 
This is specially important for particle production in mid rapidity as well as the Large Hadron Collider (LHC) 
where due to the large center of mass energy of the collision a large range of transverse momentum becomes
accessible. Furthermore, the large $x$ region will be the dominant part of kinematics covered in 
the proposed Electron Ion Collider (EIC), at least in the earliest stages~\cite{eic}.   
Therefore it is desirable to devise a formalism that not only incorporates the physics of saturation but 
also has the correct high $p_t$ and/or large $x$ physics encoded. 

Generalizing the CGC formalism to include high $p_t$ physics would have significant ramifications not
only for saturation physics and the work to determine its domain of applicability, it would also enable one
to describe a wide range of phenomena using the same formalism. For example, saturation physics is commonly
employed to provide the initial conditions for the hydrodynamic evolution of the medium, the Quark Gluon Plasma, 
created in high energy heavy ion collisions but is not applicable to jet (radiative) energy loss and the 
interactions between high $p_t$ partons and the produced medium. Elastic energy loss and loss (shift) of
rapidity are also not present in the current formulation of saturation physics but must be included in a 
more general description which could be specially significant for cold matter energy loss and 
$p_t$ broadening.
 
Toward this goal we proposed a more general approach in~\cite{jjm-elastic}. 
We considered a high energy quark scattering from a target proton or nucleus 
whereas the quark not only scatters from the small $x$ 
gluons of the target, represented by a soft color field, but also from the large 
$x$ gluons in the target. We resummed the multiple scatterings of the quark 
from the soft classical fields to all orders in the number of soft scatterings 
but kept only the first scattering from the large $x$ modes. However we 
did not consider the large $x$ modes themselves scattering from the small 
$x$ gluons. Here we continue this approach and proceed to include interactions 
between the large $x$ gluons and the soft background field representing small 
$x$ gluons of the target.

We start by a brief overview of the approximations used in high energy (eikonal) scattering and
how it is used in saturation physics when viewed in the target rest frame. We then give a brief 
summary of the approximations and the methods used in our approach in~\cite{jjm-elastic}. We 
again consider scattering of a quark from a proton or nucleus target including both small and 
large $x$ gluon modes. We then proceed to calculate and resum multiple interactions between 
the small $x$ color fields and the large $x$ gluons in the target. We then briefly outline how 
the calculated scattering amplitude may be used to extract the quark propagator in this more
general setting and how the quark propagator may be used to compute physical observables 
in full range of $x$ (and/or $p_t$).

\section{Eikonal approximation, multiple scattering at small $x$ and beyond}

Here we remind the reader of the approximations involved in high energy (eikonal) 
scattering. As this is standard and already covered in detail in~\cite{jjm-elastic} we will be 
brief here. We define the light cone coordinates as
\be
x^+ \equiv {t + z \over \sqrt{2}}\,\, , \, \, 
x^- \equiv {t - z \over \sqrt{2}}
\ee
and similarly for momenta and fields. The small $x$ gluons of the target are modeled as a soft 
color (background) field $S_a^\mu (x)$. One can either work in the frame where both 
the projectile and the target are 
moving fast, or in the frame where the projectile quark is fast and the target is at rest. In either 
frame the projectile quark moving to the right (along the $x^+$ direction to be specific) will
have a large $p^+$ component of momentum and will couple to the conjugate component of 
the target color field $S^-$. Furthermore, the target color field is independent of $x^-$ so that 
$S_a^- = S_a^- (x^+, x_t)$. We also define a light-like vector 
\be
n^\mu = (n^+ = 0, n^- = 1, n_t =0)
\label{eq:n-vector}
\ee
with $n^2 = 0$ which can be 
used to extract the Lorentz index of the soft color field and express it as 
$S^-_a \equiv n^- \, S_a (x^+, x_t)$ so that 
$\slS_a = S_s^\mu \, \gamma _\mu = \sln\, S_a$ which will help keep the 
expressions compact.
  
\begin{figure}[h]
  \centering
  \includegraphics[width=1.0\textwidth]{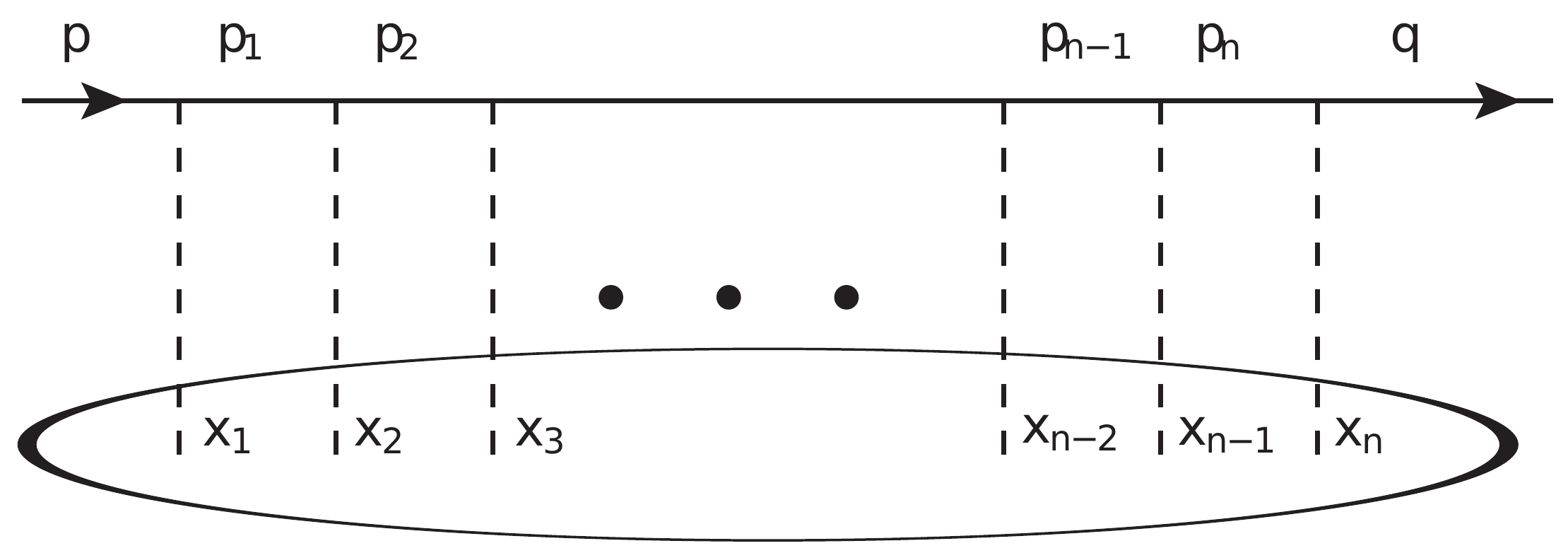}
  \caption{\it Multiple soft scatterings from the color field of the target.}
  \label{fig:n-scatt-on-target}
\end{figure}

The $n$-th order of the scattering of a quark, with momentum $p$, from the color 
field of the target is depicted in Fig. \ref{fig:n-scatt-on-target} (target is shown as an ellipse)  
where $x_i$ label the coordinate positions of the 
field $S^-$ in the target (one should think of this as the projectile quark multiply 
scattering while going right through the target so that there are no propagators 
between the quark line and the target field). Diagrams of this type resum into 
a path-ordered infinite Wilson line provided one neglects the
transverse momenta of the intermediate quark lines and the phases one picks up after
integrating over the $p^-_i$ of each intermediate quark propagator $p_i^\mu$. The 
integration over
the minus component of the intermediate propagators forces a path ordering such that
the scattering is sequential along the longitudinal direction, i.e., $x_i^+ > x_{i - 1}^+$ 
and so on (see~\cite{jjm-elastic, fg, eik-review} for details). 

\begin{figure}[h]
  \centering
  \includegraphics[width=1.0\textwidth]{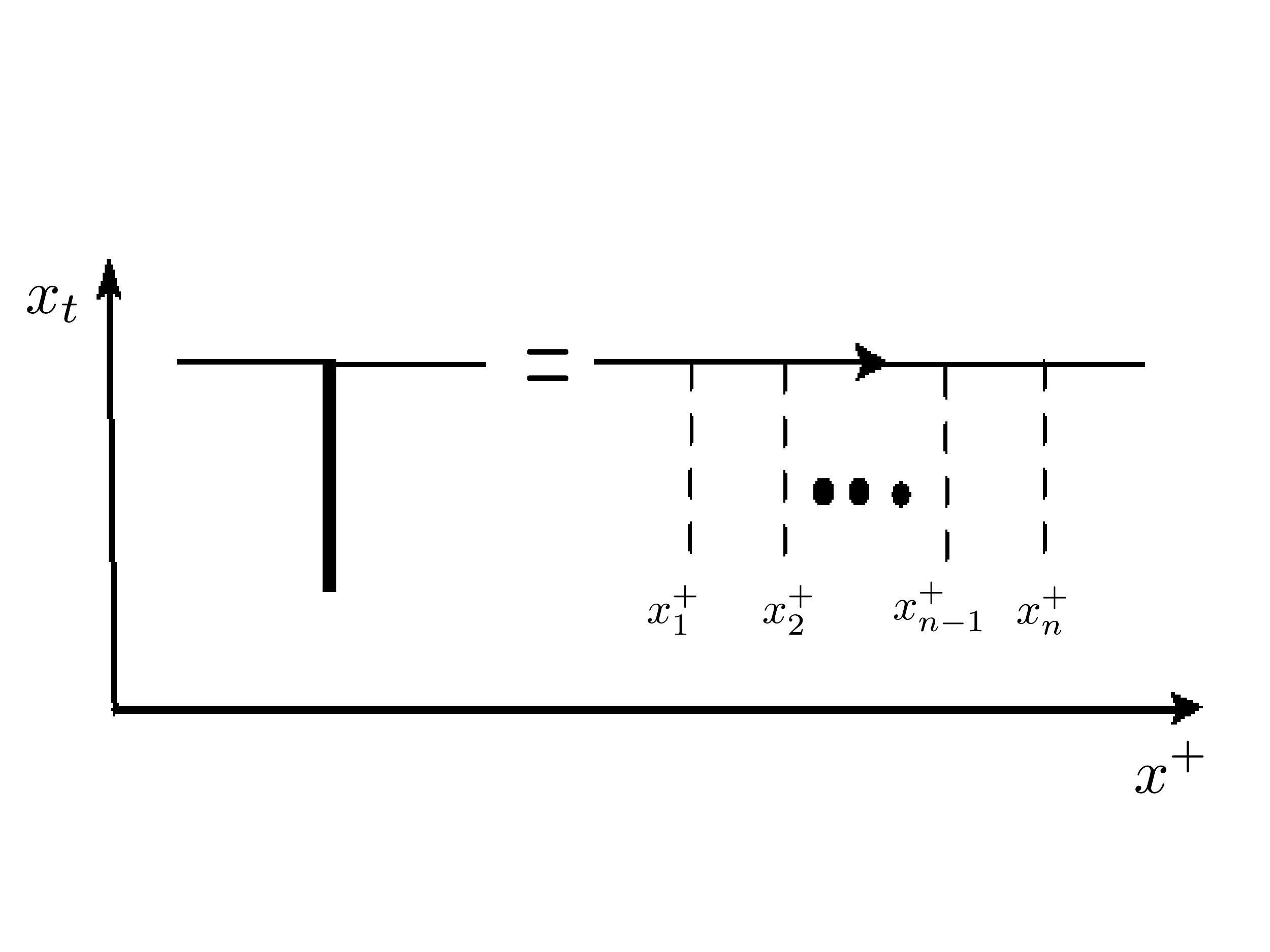}
  \caption{\it A Wilson line representing eikonal scattering of a high energy quark.}
  \label{fig:wilson-f}
\end{figure}

The scattering amplitude 
can then be written as
\be
i \mathcal{M}_{eikonal} (p,q) = 2 \pi \delta (p^+ - q^+)\,  
\ubar (q)\, \sln\, \int d^2 x_{t}\, e^{- i (q_t - p_t) \cdot x_{t}} \, 
\left[V (x_t) - 1\right]\, u(p)
\label{eq:eikonal}
\ee
where the infinite Wilson line $V (x_t)$ is defined as
\be
V (x_t) \equiv \hat{P}\, 
\exp \left\{i g \int_{- \infty}^{+\infty} d x^+ \, S^-_a (x^+, x_t)\, t_a\right\}
\ee 
and depicted in Fig.~\ref{fig:wilson-f}.
This infinite Wilson line resums multiple scatterings of a high energy quark (moving along 
the positive $z$ axis) on a soft background color field to all orders in the soft field 
$\left[i g \, S_a\, t_a\right]^n$. Due to the eikonal approximation (see also \cite{b-eik}
where the first energy suppressed terms are investigated) 
the transverse position of the quark does not change during the scattering, i.e., 
the projectile quark does not get a significant deflection (small angle scattering). 
Color matrices $t_a$ are in the fundamental representation and soft color field 
$S^-$ represents small $x$ gluon modes of the target. 

\begin{figure}[h]
  \centering
  \includegraphics[width=1.0\textwidth]{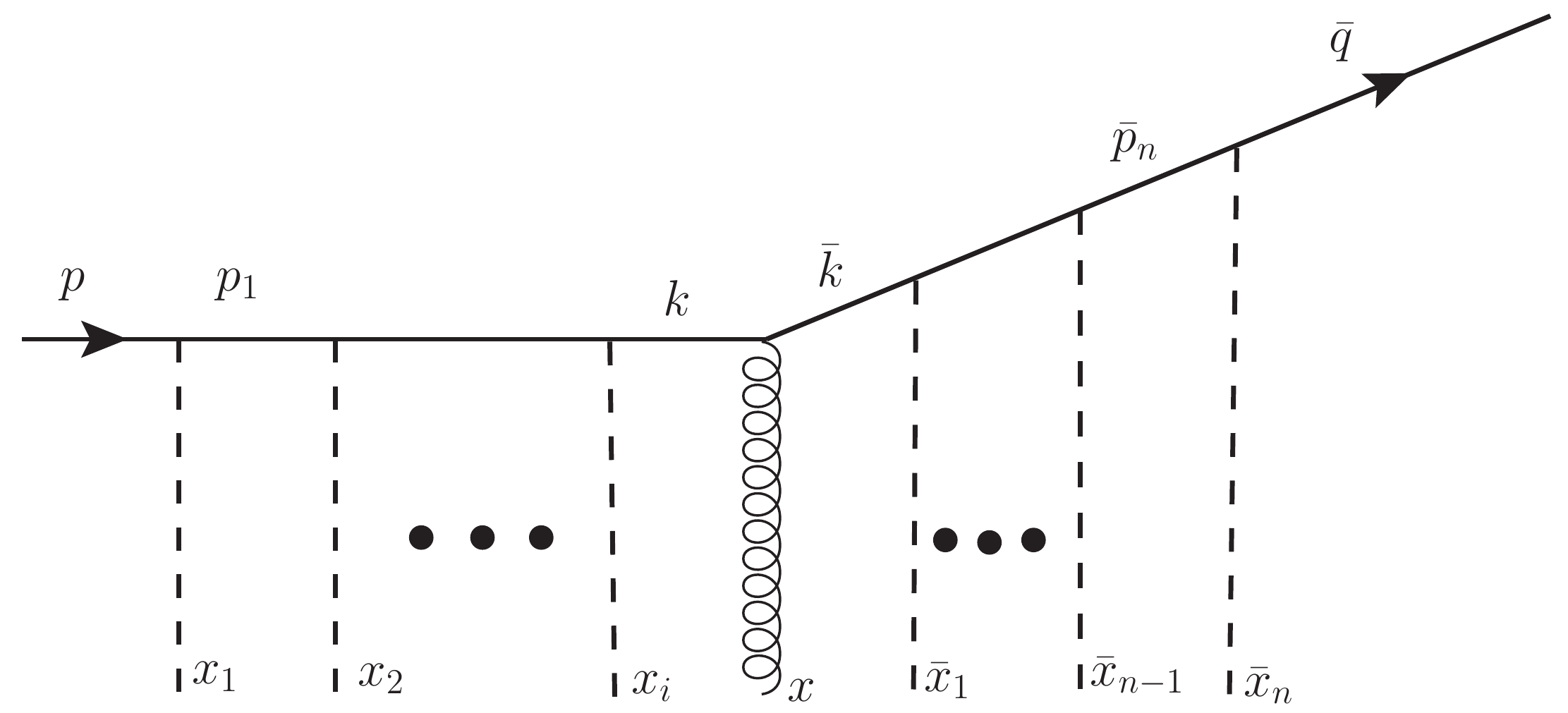}
  \caption{\it Multiple soft scatterings before and after a hard one at $x$.}
  \label{fig:nsoft-1hard-nsoft}
\end{figure}

In~\cite{jjm-elastic} we went beyond eikonal approximation by including scattering from a 
large $x$ gluon field denoted $A^\mu (x)$ which, unlike the soft field $S^-$, carries large longitudinal
momentum and can therefore cause a large deflection of the projectile quark. Due to the possibility
of this large angle deflection (so that the final state quark has a large transverse momentum) 
it was necessary to introduce a rotated frame, denoted bar-ed frame, where the scattered quark 
is moving along a new longitudinal direction $\bar{z}$. The bar-ed coordinates are related to 
the original  $x,y,z$ coordinates (projectile quark is moving along the $z$ direction) via the rotation 
matrix $\mathcal{O}$ in $3$ dimensions.
\be
\left(\begin{array}{c}
\bar{x} \\
\bar{y} \\
\bar{z}
\end{array} \right)
 = 
\mathcal{O} \,
\left(\begin{array}{c}
x  \\
y \\
z 
\end{array} \right)
\ee
The elements of this $3$-d rotation matrix are expressed in terms of the $3$-momentum of the 
scattered quark. We also defined a new light cone vector $\bar{n}$ which projects out the plus 
component in the new frame, so that $\bar{n}\cdot \bar{p} = \bar{p}^+$. 
We then managed to resum all the multiple scatterings of the projectile quark from the soft field 
and one scattering from the large $x$ (sometimes referred to as the hard field where hard refers to
a large longitudinal momentum). This is shown in Fig.~\ref{fig:nsoft-1hard-nsoft} (target is not 
explicitly drawn),

Diagrams of this type resum into~\cite{jjm-elastic} 
\bea
i \mathcal{M}_1 &=&  \int d^4 x\, d^2 z_t \, d^2 \bar{z}_t \, 
\int {d^2 k_t \over (2 \pi)^2} \, {d^2 \bar{k}_t \over (2 \pi)^2} \, 
e^{i (\bar{k} - k) x} \, 
e^{- i (\bar{q}_t - \bar{k}_t)\cdot \bar{z}_t}\, 
e^{- i (k_t - p_t)\cdot z_t}
 \nn
&&
\ubar (\bar{q})\, \left[ 
\overline{V}_{AP} (x^+, \bar{z}_t) \, \slnbar \, {\slkbar \over 2 \bar{k}^+} \,
\left[ i g \, \slA (x)\right]\, 
{\slk \over 2 k^+} \, \sln \, V_{AP} (z_t, x^+)  
\right]\, u(p)
\label{eq:nsoft-hard-nsoft}
\eea
with $k^+ = p^+, k^- = {k_t^2 \over 2 k^+}$, 
$\bar{k}^+ = \bar{q}^+, \bar{k}^- = {\bar{k}_t^2 \over 2 \bar{k}^+}$ 
and the semi-infinite, anti path-ordered Wilson lines in the fundamental representation are now defined 
as~\footnote{See \cite{jjm-elastic} regarding the use of path vs. anti path-ordered 
Wilson lines in propagators vs. amplitudes.}    
\be
\overline{V}_{AP} (x^+, \bar{z}_t) \equiv \hat{P}\, 
\exp \left\{i g \int_{x^+}^{+\infty} d \bar{z}^+ \, \bar{S}^-_a 
(\bar{z}_t, \bar{z}^+)\, t_a\right\}
\label{eq:Wilsonbar-si-fundamental}
\ee
and
\be
V_{AP} (z_t, x^+) \equiv \hat{P}\, 
\exp \left\{i g \int_{- \infty}^{x^+} d z^+ \, S^-_a (z_t, z^+)\, t_a\right\}.
\label{eq:Wilson-si-fundamental}
\ee 
where anti path-ordering (AP) in the amplitude means fields with the largest argument appear to the left.   
\subsection{Multiple scatterings of the large $x$ gluon}

\begin{figure}[h]
  \centering
  \includegraphics[width=0.75\textwidth]{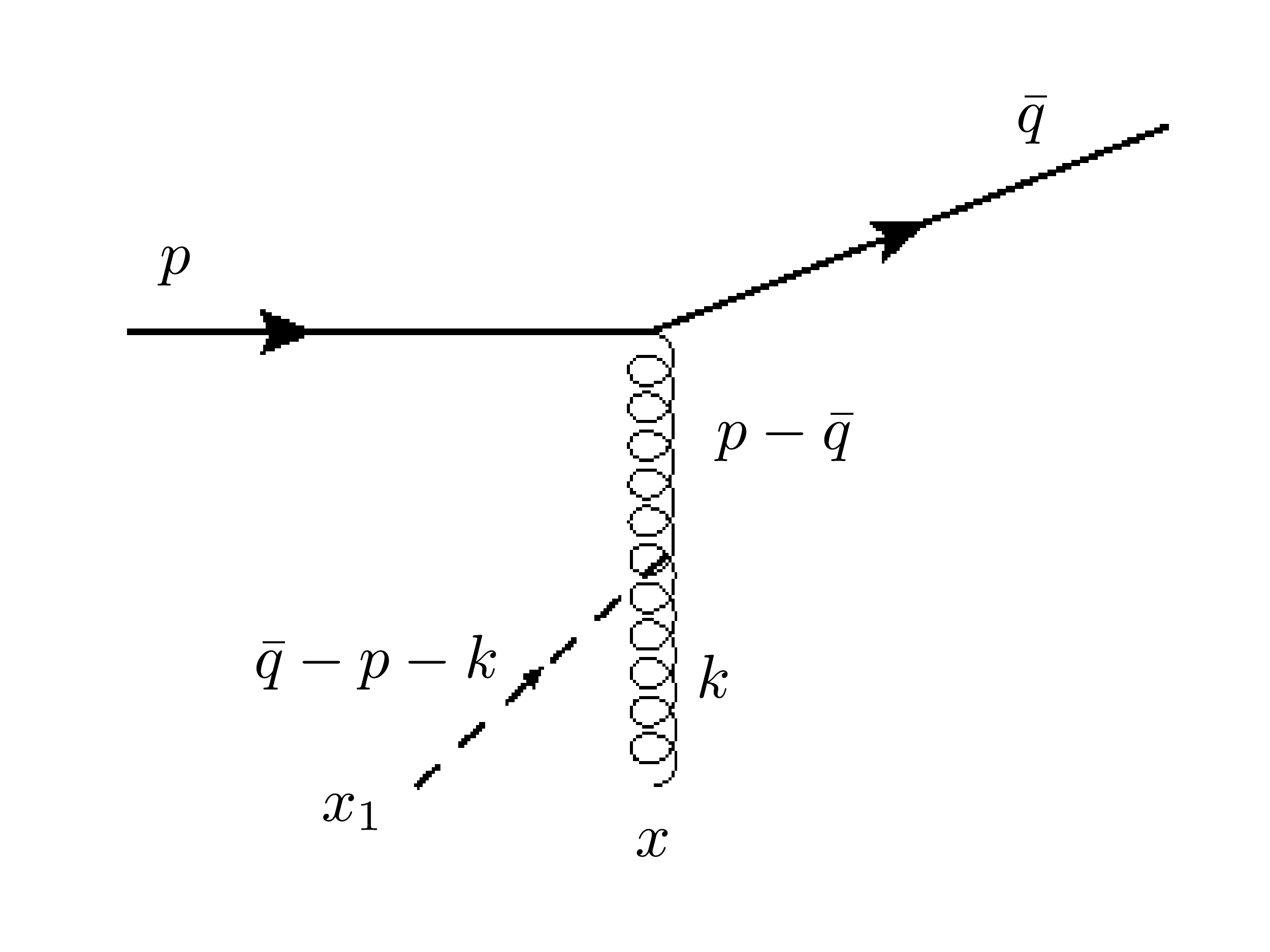}
  \caption{\it One soft scattering of the large $x$ gluon.}
  \label{fig:1soft-on-hard}
\end{figure}

\begin{figure}[h]
  \centering
  \includegraphics[width=0.75\textwidth]{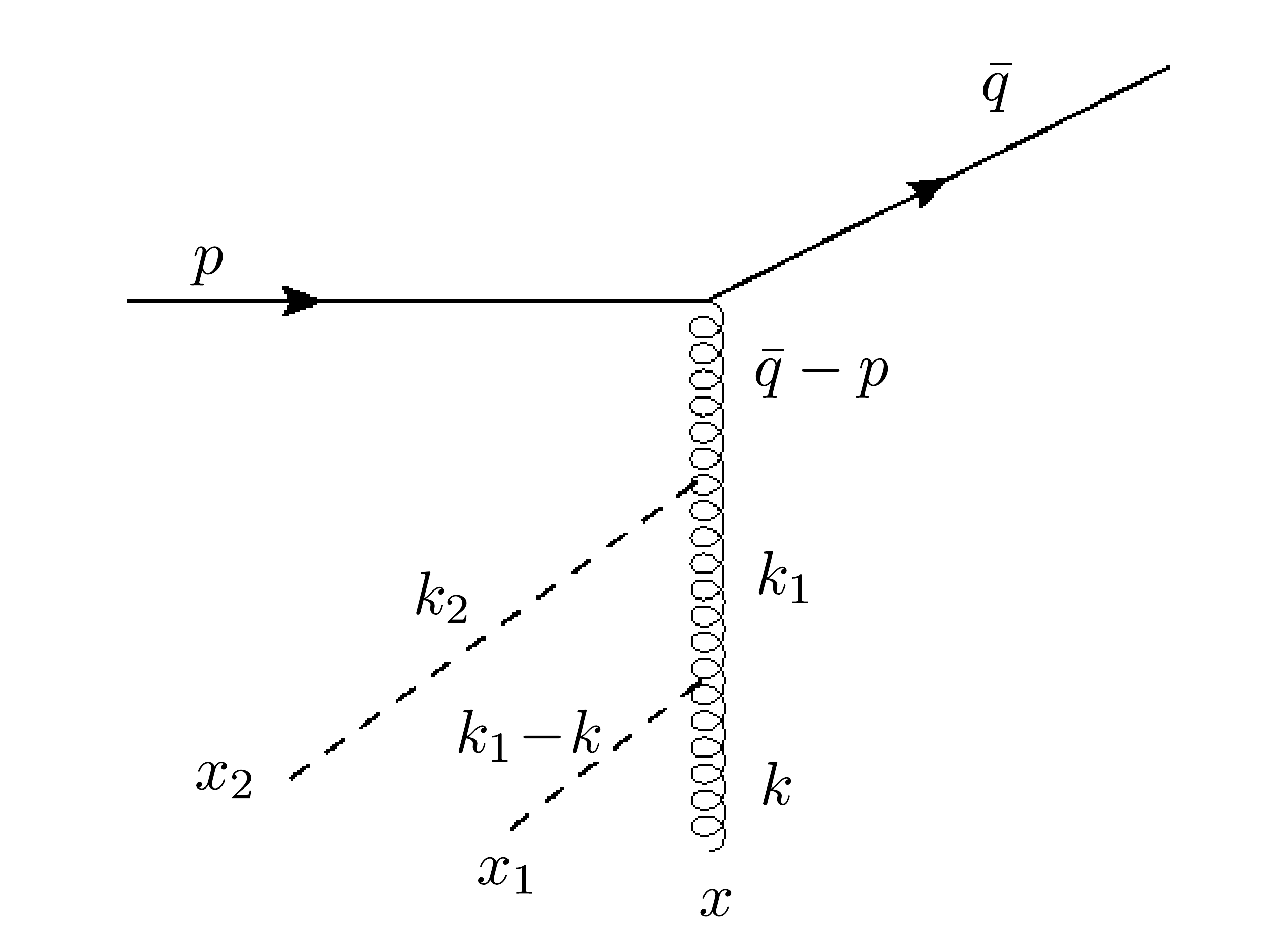}
  \caption{\it Two soft scatterings of the large $x$ gluon.}
  \label{fig:2soft-on-hard}
\end{figure}

\begin{figure}[h]
  \centering
  \includegraphics[width=0.75\textwidth]{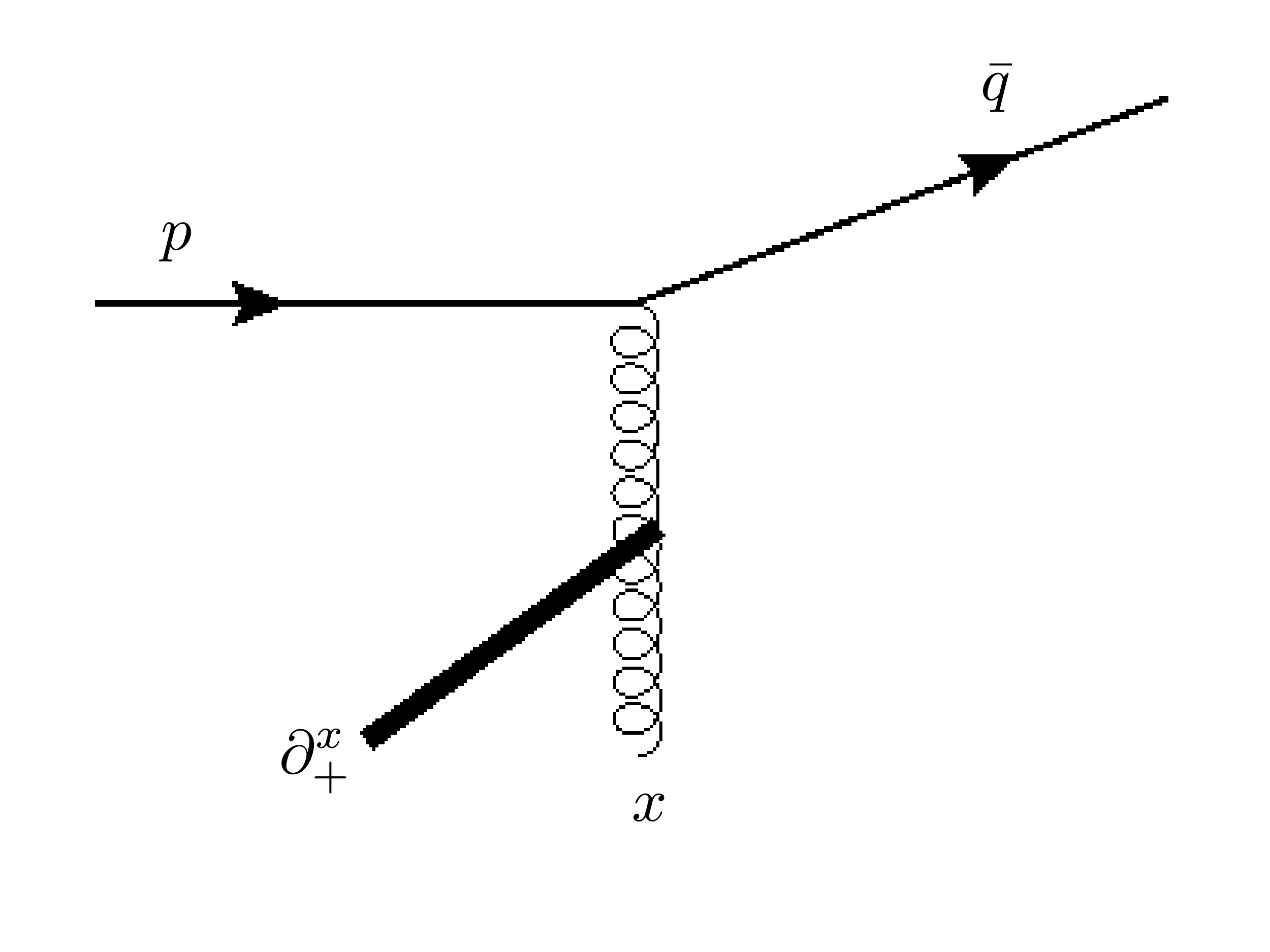}
  \caption{\it Multiple soft scatterings of the large $x$ gluon.}
  \label{fig:nsoft-on-hard}
\end{figure}

We now proceed to consider interactions of the hard (large $x$) gluon with the soft background field. First
let us consider the case when only the hard gluon interacts with the soft field but not the initial or 
final state quark, as shown in Fig.~\ref{fig:1soft-on-hard}
The amplitude can be written as 
\bea
i \mathcal{M} &=&  \int {d^4 k \over (2 \pi)^4}\, d^4 x \, d^4 x_1 \, e^{i (\bar{q} - p -k) x_1}\, e^{i k x} \,
\bar{u} (\bar{q}) \, (i g\, \gamma^\mu \, t^a)\, u (p) \nn 
&& G_{\mu\nu}^{a b} (p - \bar{q}) \, 
V^{\nu \, \lambda\rho}_{b c d} (p - \bar{q}, k, \bar{q} - p -k) 
 A_\lambda^c (x) \, \left[i\, g \, S_\rho^d (x_1)\right]\nonumber 
\eea
where the free gluon propagator is 
\be
G^{\mu\nu}_{a b} (l) = {i \delta_{a b} \over l^2 + i \epsilon} \, 
\left[ - g^{\mu\nu} + {n^\mu\, l^\nu + l^\mu\, n^\nu \over n\cdot l} \right]
\ee
and the triple gluon vertex is
\be
V^{b c d}_{\nu\lambda\rho} (l_1, l_2, l_3) = g \, f^{b c d}\, 
\left[(l_2 - l_3)_\nu \, g_{\lambda \rho} + (l_3 - l_1)_\lambda\, g_{\rho \nu} 
+ (l_1 - l_2)_\rho \, g_{\nu \lambda} \right]
\ee
The large $x$ gluon field is denoted $A (x^+, x^-, x_t)$ while the soft field is 
$S^-_d (x_1^+, x_{1 t}) = n^- \, S_d (x_1^+, x_{1 t})$. 
It is straightforward to simplify the Lorentz structure of the amplitude by repeated use of 
the gauge condition $n\cdot A = 0$ as well as the null condition $n^2 = 0$ and 
the fact that the soft field has only a $-$ component and therefore is 
proportional to $n^\mu$. We get
\bea
i \mathcal{M} &=&  f_{a c d}\, \int {d^4 k \over (2 \pi)^4}\, d^4 x \, 
d^4 x_1 \, e^{i (\bar{q} - p -k) x_1}\, e^{i k x} \,
\bar{u} (\bar{q}) \, (i g\, \gamma^\mu \, t^a)\, u (p) \, 
A_\lambda^c (x) \, \left[i\, g \, S^d (x_1)\right]\nn
&&
{1 \over (p - \bar{q})^2 + i \epsilon} \, 
\left[- g^\mu_\lambda \, n\cdot (p - \bar{q} - k) + n^\mu \, (p - \bar{q}_\lambda\,
(1 - {n \cdot k \over n\cdot (p - \bar{q})})\right]
\eea
As before the soft field $S$ is independent of the $x_1^-$ coordinate which allows us to do
the integration over $x_1^-$ coordinate which gives $\delta (\bar{q}^+ - p^+ - k^+)$
which in turn is used to do the $k^+$ integration setting $k^+ = \bar{q}^+ - p^+$ upon 
integrating $k^+$. We also note
that the only $k$ dependence left is in the phase factors since $n \cdot k$ kills all the 
other possible $k$ dependent terms. Then one can carry out the $k_t$ and $k^-$ integrations 
which lead to delta functions $\delta^2 (x_{1 t} - x_t)\, \delta (x_1^+ - x^+)$ which 
are then used to perform the remaining integrations over $x_1^+, x_{1 t}$ 
setting $x_{1 t} = x_t$ and $x_1^+ = x^+$. In other words the soft field $S$ and the 
large $x$ (hard) field $A$ are at the same space coordinate except that the soft field $S$ 
does not depend on $x^-$, unlike the hard field $A$. After performing the above integrations 
we are left with
\bea
i \mathcal{M} &=& 2 f_{a c d}\, \int \, d^4 x \,  e^{i (\bar{q} - p) x}\, 
\bar{u} (\bar{q}) \, 
{\left[\sln \, (p - \bar{q}) \cdot A_c (x) - \slA_c (x) \, n\cdot (p - \bar{q})\right]\over
(p - \bar{q})^2}\, 
(i g\, t^a)\, 
u (p)  \nn
&&
 \left[i\, g \, S^d (x^+, x_t)\right]
\eea
with the most important point being the soft and hard field are at the same point. We now go ahead and
consider one more soft scattering of the hard gluon as shown in Fig.~\ref{fig:2soft-on-hard},
We can now repeat the same steps as before, using the gauge choice and the null vector condition
to simplify the Lorentz structure, performing the integration over $k$ puts the fields $A$ and
$S (x_1)$ at the same point ($x_1 = x$) as before, however there is now an integration 
over $k_1$. Let us now consider the $k_1^-$ integration,
\be
I \equiv \int {d k^-_1 \over (2 \pi)} \, {e^{i k_1^- (x^+ - x_2^+)} \over 
2 (\bar{q}^+ - p^+) \left[ k_1^- - {k_{1 t}^2 - i \epsilon \over 2 (\bar{q}^+ - p^+)}\right]}
\ee
This integration can be done using the standard contour integration techniques realizing that the
$k_1^-$ pole is always below the real axis since $p^+ > \bar{q}^+$. This integral is then  
\be
I \sim {i\, \theta (x^+ - x_2^+) \over 2 n \cdot (\bar{q} - p)}\, 
e^{i {k_{1 t}^2 \over 2 n\cdot (\bar{q} -p)} (x^+ - x_2^+)}\, .
\ee
The most essential point here is the reappearance of the theta function $\theta (x^+ - x_2^+)$ which
forces a path ordering of the soft scatterings. This is not surprising since the soft multiple scatterings
are eikonal. The rest of the analysis goes through as before and we get
\bea
i \mathcal{M} &=& 2  \, f_{a b c }\, f_{c d e} \int \, d^4 x \,  d x_2^+\, \theta (x^+ - x_2^+)\, 
e^{i (\bar{q} - p) x}\nn
&&
\bar{u} (\bar{q}) \, 
{\left[\sln \, (p - \bar{q}) \cdot A_e (x) - \slA_c (x) \, n\cdot (p - \bar{q})\right]\over
(p - \bar{q})^2}\, 
(i g\, t^a)\, 
u (p) \nn
&& \left[i\, g \, S_d (x^+, x_t)\right]\, \left[ i\, g \, S_b (x_2^+, x_t)\right]
\eea
Including more soft scattering on the hard gluon line is straightforward and proceeds as usual, with 
each extra soft scattering path ordered. This allows one to resum all the soft scatterings of the hard gluon 
and write it as
\bea
i \mathcal{M}_2 &=& {2\, i \over (p - \bar{q})^2} \, \int d^4 x\, 
e^{i (\bar{q} - p) x} \, 
\ubar (\bar{q})\, \bigg[ (i g \, t^a)\, 
\left[\partial_{x^+}\, U_{AP}^\dagger (x_t, x^+)\right]^{a b} \nonumber\\
&&
\left[n \cdot (p - \bar{q}) \, \slA_b (x) -  (p - \bar{q}) \cdot A_b (x) \, \sln\right]\, 
\bigg]\, u(p)
\label{eq:nsoft-on-hard}
\eea
where the derivative acts on the $+$ coordinate of the adjoint Wilson line (anti path-ordered in the amplitude) 
and arises from the fact that
one can write the soft field at the last scattering point as a derivative on the Wilson line. 
This amplitude is symbolically shown in 
Fig.~\ref{fig:nsoft-on-hard} where the thick solid line attached to the hard gluon line depicts a 
semi-infinite (and anti path-ordered) Wilson line in the adjoint representation 
(analog of eq. (\ref{eq:Wilson-si-fundamental}))~\footnote{Recall that Adjoint representation is real so
that $\left[U^\dagger\right]^{a b} = U^{b a}$.}. 
Finally we note that this amplitude vanishes in the soft (eikonal) limit, i.e. when $\bar{q}^+ \rightarrow p^+$
and $A^\mu \rightarrow S^-$.

\subsection{Multiple scatterings of the large $x$ gluon and the final state quark}
\begin{figure}[h]
  \centering
  \includegraphics[width=1.0\textwidth]{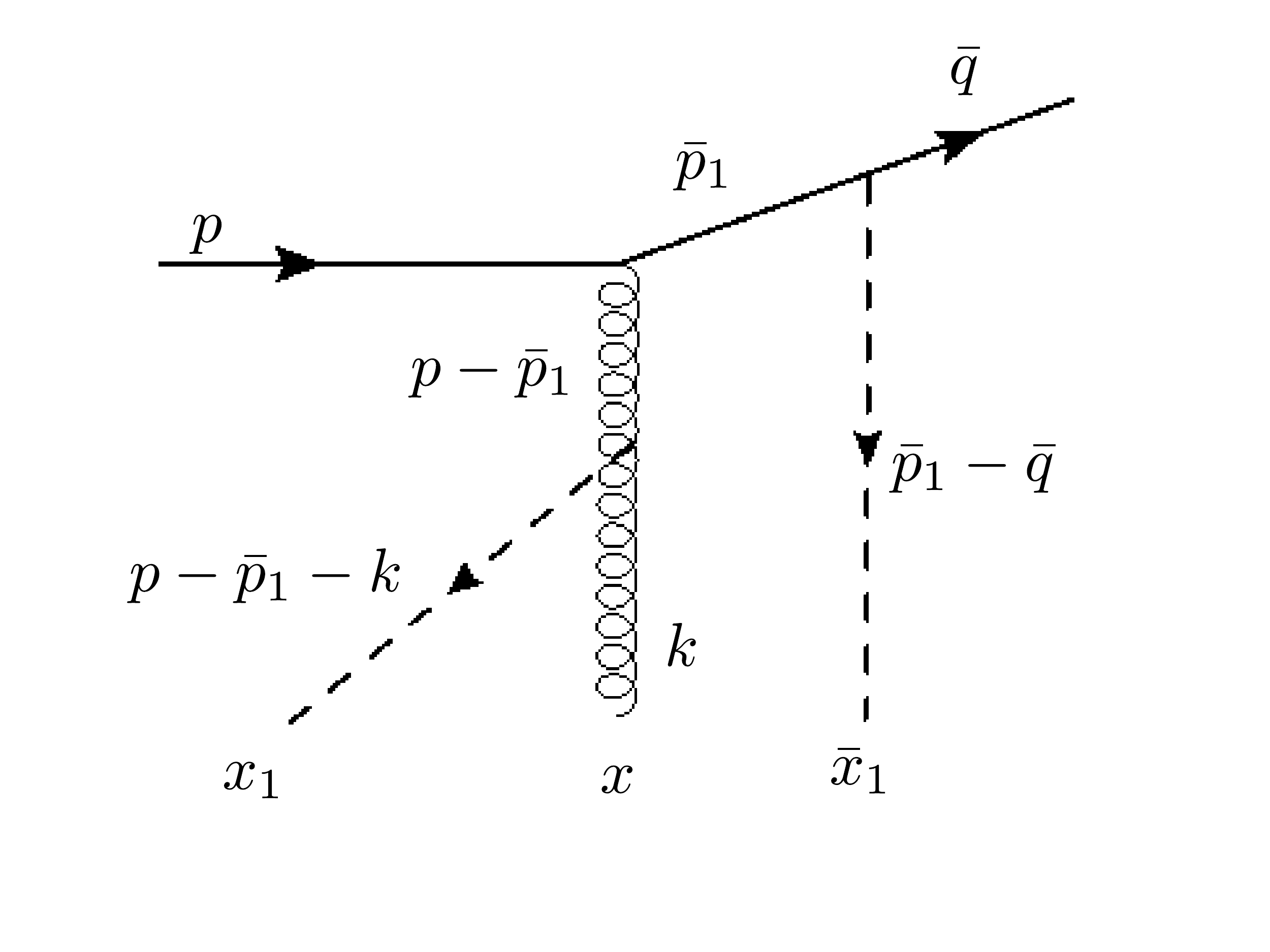}
  \caption{\it Both the large $x$ gluon and the final state quark scattering once.}
  \label{fig:1soft-on-hard-1soft-on-final}
\end{figure}

\begin{figure}[h]
  \centering
  \includegraphics[width=0.75\textwidth]{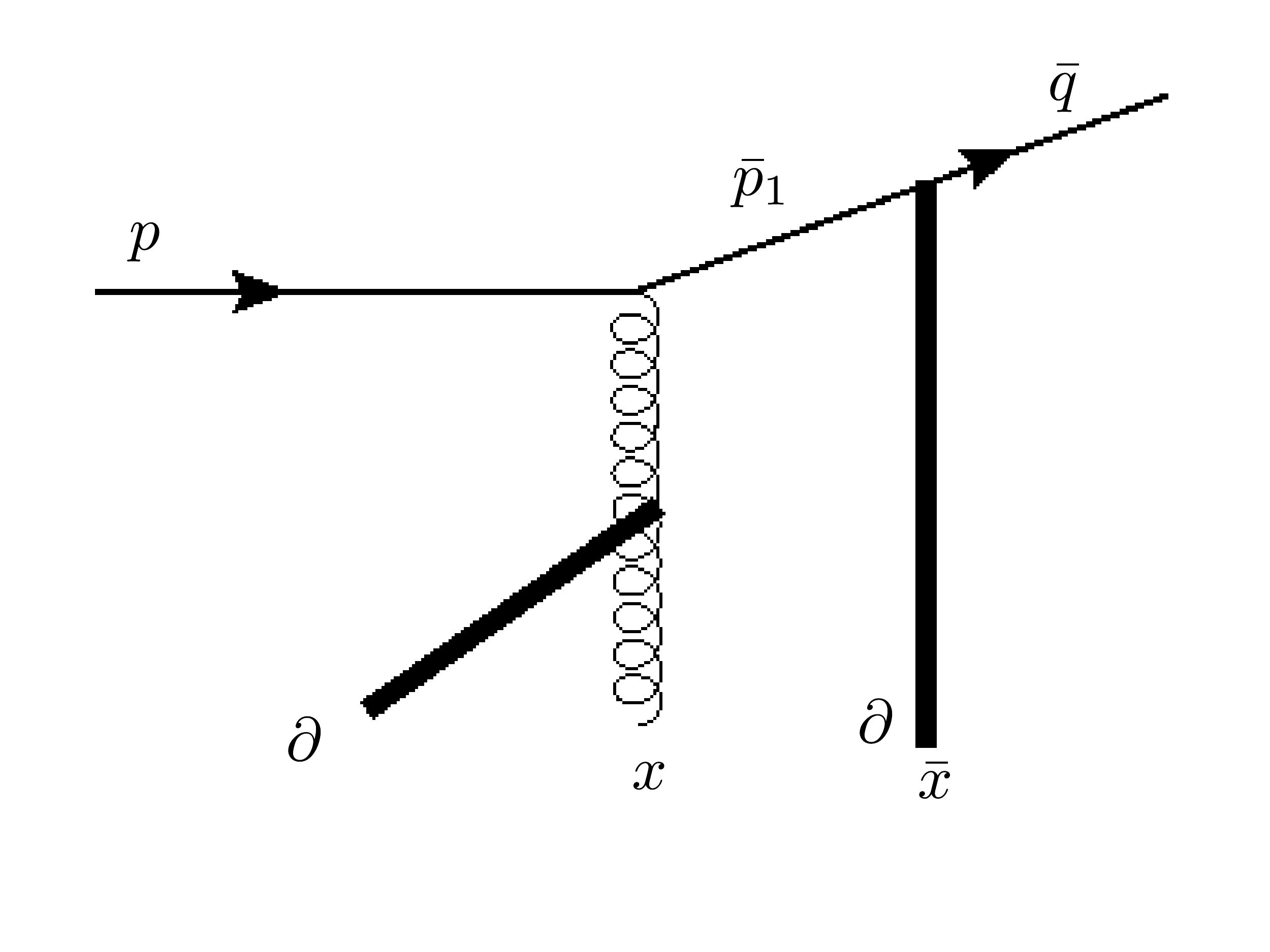}
  \caption{\it Soft scatterings of the final state quark and the large $x$ gluon.}
  \label{fig:nsoft-on-hard-nsoft-on-final}
\end{figure}

We now consider the case when both the large $x$ gluon and the final state quark
multiply scatter from the soft background field. The first diagram not included so 
far is when both the large $x$ gluon and the final state quark scatter once as shown
in Fig.~\ref{fig:1soft-on-hard-1soft-on-final}. 
This scattering amplitude is given by
\bea
i \mathcal{M} &=&  \int {d^4 k \over (2 \pi)^4}\, {d^4 \bar{p}_1 \over (2 \pi)^4}\,
d^4 x \, d^4 x_1 \, d^4 \bar{x}_1 \, e^{i (\bar{q} - \bar{p}_1) \bar{x}_1}\, e^{- i k x} \,
e^{- i (p - \bar{p}_1 - k) x_1}\, \bar{u} (\bar{q}) \, \slnbar \nn
&&
\left[i\, g \, \bar{S} (\bar{x}_1)\right] \, {i \bar{\slp}_1 \over \bar{p}_1^2 + i \epsilon}
(i g\, \gamma^\mu \, t^a)\, u (p) \, G_{\mu\nu} (p - \bar{p}_1) \, 
V^{\nu\lambda\rho}_{a b c} (p - \bar{p}_1, - k, \bar{p}_1 + k - p) \nn
&&
 A_\lambda^b (x) \, \left[i\, g \, n_\rho \, S^c (x_1)\right]\nonumber 
\eea
Many of the steps involved in simplifying this expression are identical to the previous ones, 
for example we use the fact that the soft field $S^- (x_i)$ is independent of the $x_i^-$ 
coordinate to perform the integration over the minus
components of their coordinates leading to delta functions relating the $+$ components of the momenta.
The Lorentz structure can also be simplifies as before by repeated use of the gauge condition as well
as the fact that $n^2 = \bar{n}^2 = 0$ and that the soft fields (in their respective frames) have only
minus components which allows extraction of their Lorentz index by the use of the light-like vectors $n$ and
$\bar{n}$ so that $S^- = n^- \, S$ and $\bar{S}^- = \bar{n}^- \, \bar{S}$. As in the case of only
the hard gluon scattering we just considered one gets $x_1^+ = x^+$ and $x_{1 t} = x_t$. The 
most important part is the integration over $\bar{p}^-_1$ momentum of the intermediate quark
line which can be written as
\be
I \sim \int {d \bar{p}^-_1 \over (2 \pi)} \, {e^{ - i \bar{p}_1^- ( \bar{x}_1^+ - x^+)} \over
\left[\bar{p}_1^- - {\bar{p}_{1 t}^2 - i \epsilon \over 2 \bar{q}^+} \right]
\left[\bar{p}_1^- - p^- -{(\bar{p}_{1 t} - p_t)^2 - i \epsilon \over 2 (\bar{q}^+ - p^+)} \right]
}
\ee
keeping in mind that both $p^+, \bar{q}^+ > 0$ and that $p^+ - \bar{q}^+ > 0$ we see that
the integral above has two poles which are on the opposite side of the real axis. This is completely
different from the eikonal scattering where the intermediate quark propagators have poles which
are all on the same side of the real axis which leads to path ordering along the $+$ direction. 
This integral can be evaluated using the standard contour integration techniques and gives 
\be
I \sim \left[
\theta (\bar{x}_1^+ - x^+)\,  e^{ - i {\bar{p}_{1 t}^2 \over 2 \bar{q}^+} ( \bar{x}_1^+ - x^+)} 
+ \theta (x^+ - \bar{x}_1^+)\, 
e^{ - i \left[p^- + {(\bar{p}_{1 t} - p_t)^2 \over 2 (\bar{q}^+ - p^+)}\right] 
(\bar{x}_1^+ - x^+) }\right]
\ee
To proceed further and to stay consistent with the approximations made for strict eikonal 
scattering where one neglects terms of the order ${p_t \over p^+}$ we will ignore the 
phase factors above. We then see that the two different path orderings corresponding to the
two theta functions add to unity and path ordering disappears. This can be understood 
as the following, integration over any of the poles forces the other propagator to go off-shell 
and to become space-like in which case there is no absolute ordering between the interaction 
vertices at $x_1^+ = x^+$ and $\bar{x}_1^+$. However, if we consider further soft 
scatterings of the hard gluon and the final state quark they will be path ordered with respect to
$x^+$ and  $\bar{x}_1^+$ respectively. This is straightforward but long and we will just 
quote the final result as the calculations proceeds as earlier and there are no further subtle 
points. Resuming all the soft scatterings of the hard gluon and the final state quark then gives
\bea
i \mathcal{M}_3 &\!\!=\!\!& - 2\, i\, \int d^4 x\,  d^2 \bar{x}_t \, d \bar{x}^+\,  
 {d^2 \bar{p}_{1 t} \over (2 \pi)^2} \, 
e^{i (\bar{q}^+ - p^+) x^-} \, 
e^{- i (\bar{p}_{1 t} - p_t)\cdot x_t}\, 
e^{- i (\bar{q}_t - \bar{p}_{1 t})\cdot \bar{x}_t}
 \nn
&&
\ubar (\bar{q})\, \bigg[
\left[\partial_{\bar{x}^+}\, \overline{V}_{AP} (\bar{x}^+, \bar{x}_t)\right]\, 
 \slnbar\, \slpbarone \,
(i g t^a)\, 
\left[\partial_{x^+}\, U^\dagger_{AP} (x_t, x^+)\right]^{a b}\nn
&&
{
\left[ n \cdot (p - \bar{q}) \, \slA^b (x) -  (p - \bar{p}_1) \cdot A^b (x) \, \sln\right]
\over
\left[
2 n \cdot \bar{q} \, 2 n \cdot (p - \bar{q})\, p^- 
- 2 n \cdot (p - \bar{q})\, \bar{p}^2_{1 t} 
- 2 n \cdot \bar{q} \, (\bar{p}_{1 t} - p_t)^2
\right]
}
\bigg]\, u (p)
\label{eq:nsoft-on-hard-nsoft-on-final}
\eea
and is depicted in Fig.~\ref{fig:nsoft-on-hard-nsoft-on-final} below where the thick solid
lines denote semi-infinite (and anti path-ordered) Wilson lines in fundamental 
(attached to the final state quark line) and adjoint (attached to the hard gluon line) 
representations. Also, we have $\bar{p}_1^+ = \bar{q}^+$. We again note that this
amplitude also vanishes in the soft limit.

\subsection{Multiple scatterings of the initial state quark and the large $x$ gluon}

\begin{figure}[h]
  \centering
  \includegraphics[width=0.75\textwidth]{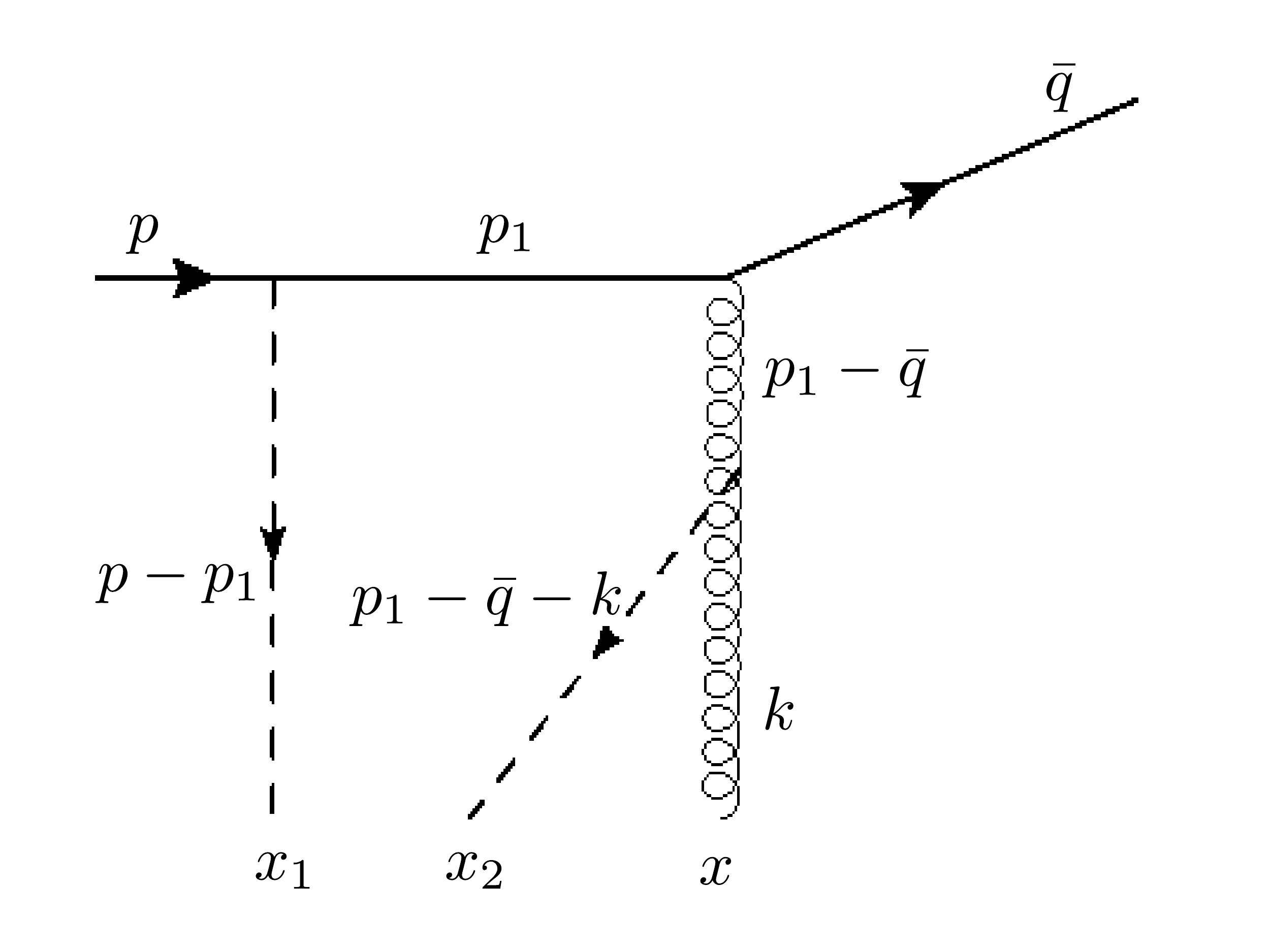}
  \caption{\it Soft scatterings of the initial state quark and the large $x$ gluon.}
  \label{fig:1soft-on-initial-1soft-on-hard}
\end{figure}

We now consider the next class of diagrams in which both the large $x$ gluon and the
initial state quark scatter from the soft background field. The lowest order diagram, 
not included so far, is shown in Fig.~\ref{fig:1soft-on-initial-1soft-on-hard} below
This amplitude can be written as 
\bea
i \mathcal{M} &=&  \int {d^4 k \over (2 \pi)^4}\, {d^4 p_1 \over (2 \pi)^4}\,
d^4 x \, d^4 x_1 \, d^4 x_2 \, e^{i (p_1 - p) x_1}\, e^{- i k x} \,
e^{ i (k + \bar{q} - p_1) x_2}\nn
&&
\left[ \bar{u} (\bar{q})  \, (i g\, \gamma^\mu \, t^a) \, 
{i \slp_1 \over p_1^2 + i \epsilon}\, \sln \, \left[i\, g \, S (x_1)\right] \, u (p) \right]
\, G_{\mu\nu} (p_1 - \bar{q}) \nn
&&
V^{\nu\lambda\rho}_{a c d} (p_1 - \bar{q}, - k, k+ \bar{q} - p_1) \, 
 A_\lambda^c (x) \, \left[i\, g \, n_\rho \, S^d (x_2)\right]\nonumber 
\eea
Again most of the steps are identical to before; one integrates over $-$ component of the soft fields
coordinates, eventually setting some $+$ components of momenta equal to each other. 
Lorentz structure is simplified using the gauge condition and the null vector $n^2=0$ as well
as extracting the Lorentz index of the soft field $S^- = n^\mu \, S$. Integration over $k$ 
again sets $x^+_2 = x^+$. The crucial step now is the integration over $p_1^-$ which we
focus on,
\be
I \sim \int {d p^-_1 \over (2 \pi)} \, {e^{ - i p_1^- ( x_1^+ - x^+)} \over
\left[p_1^- - {p_{1 t}^2 - i \epsilon \over 2 p^+} \right]
\left[p_1^- - \bar{q}^- -{(p_{1 t} - \bar{q}_t)^2 - i \epsilon \over 2 (p^+ - \bar{q}^+)} \right]
}
\ee
The two poles are now both below the real axis so that the integration gives a non-zero value only
if $x^+ > x_1^+$, we get
\be
I \sim \theta (x^+ - x_1^+) \, 
\left[
e^{  i {p_{1 t}^2 \over 2 p^+} ( x_1^+ - x^+)} 
- 
\, e^{  i \left[\bar{q}^- + {(p_{1 t} - \bar{q}_t)^2 \over 2 (p^+ - \bar{q}^+)}\right] 
(x_1^+ - x^+) }\right]
\ee
now, unlike in the case of scattering from the large $x$ gluon and the final state quark, the
relative sign between the two phase factors is negative (recall that the poles were on the
opposite side in that case). Hence ignoring the phases again consistent with strict eikonal 
approximation, one gets a cancellation between the two terms so that $I = 0$, therefore 
this amplitude identically vanishes! It is 
straightforward to include any number of further soft scatterings from the initial or final state quark
or the large $x$ gluon. However, it can be shown these further scatterings do not affect this null 
result. Therefore we conclude that one can not have simultaneous soft scatterings from the initial 
state quark and the large $x$ gluon. There are no other diagrams to consider, therefore this completes 
our derivation of the amplitude for scattering of a quark from the small and large $x$ gluon fields 
of the target. 

The result for the full scattering amplitude at all $x$ (or any $p_t$) can thus be written as
\be 
 i \mathcal{M} = i \mathcal{M}_{eikonal} + i \mathcal{M}_1 + 
i \mathcal{M}_2 +  i \mathcal{M}_3
\ee
where $i \mathcal{M}_{eikonal}$,  $i \mathcal{M}_1$,  
$i \mathcal{M}_2$ and  $i \mathcal{M}_3$ are given by 
eqs.~(\ref{eq:eikonal},\ref{eq:nsoft-hard-nsoft},\ref{eq:nsoft-on-hard},\ref{eq:nsoft-on-hard-nsoft-on-final}) 
respectively. This is our main result. We further recall~\cite{jjm-elastic} 
that one needs to use the 
covariant coupling in the large $x$ terms so that the above amplitude 
smoothly reduces to the eikonal amplitude $i \mathcal{M}_{eikonal}$ 
in the soft scattering limit, i.e. when 
$\slA (x^+, x^-, x_t)\rightarrow \sln\, S(x^+, x_t)$. To extract
the quark propagator one defines the effective vertex $\tau_F$ as
\be
i \mathcal{M} (p,\bar{q}) = \ubar (\bar{q})\, \tau_F (p,\bar{q})\, u (p)
\ee
in terms of which the propagator can be written as  
\be
\label{eq:prop-gen-both}
S_F (p,\bar{q}) = (2 \pi)^4 \delta^4 (p - \bar{q})\, S_F^0 (p) +  
S_F^0 (p)\, \tau_F (p,\bar{q}) \, S_F^0 (\bar{q})
\ee

A final remark is in order here, we have treated the target as consisting 
of gluons only and have totally ignored the contribution of quarks at large $x$. 
We expect sea quarks to appear when one performs a one loop correction to 
our result and can therefore be included in principle. On the other hand 
inclusion of valence quarks at large $x$ is an open problem at this point and 
will require a detailed investigation which is beyond the scope of this work.

\section{Discussion and summary}

We have derived the amplitude for scattering of a high energy quark on the gluon field of 
a proton or nucleus target including both small and large $x$ gluon modes of the target. 
This generalizes the standard expressions for eikonal scattering and is, to the best of our 
knowledge, the first gluon saturation-based calculation which includes large $x$ gluons (in
the target proton or nucleus) as dynamical degrees of freedom. As such it allows one to 
investigate many important high $p_t$ and/or large $x$ phenomena which are not 
accessible to the standard gluon saturation formalism. This is specially essential for 
a proper and quantitative understanding of the outcome of the experiments at the  
proposed Electron Ion Collider and the Large Hadron Collider.

The derived scattering amplitude is a "tree-level" expression which can already be used to investigate several
phenomena, for example $p_t$ broadening and elastic energy loss as well as the nuclear modification 
factor $R_{pA}$~\cite{jjm-rpA}. Due to the presence of large $x$
gluon fields in the target the scattered quark can undergo an arbitrarily large deflection and 
pick up large transverse momenta. Furthermore, it can lose longitudinal momentum and 
undergo a potentially large rapidity loss which is
not contained in saturation formalism. One can also extract the quark propagator from this  
scattering amplitude and use it to calculate the tree-level production cross sections for particle production
in high energy collisions for any transverse momentum $p_t$ (at small or large $x$). One would also
need to calculate the gluon propagator~\cite{gluon-prop} in this formalism which is a 
straightforward extension of the present work. This would allow one to investigate cold matter
radiative energy loss including both the fully coherent, present in saturation formalism, as well as the 
partially coherent/incoherent energy loss which is not present in the saturation formalism. 
Examples of where our results in the present form 
can be used are single inclusive or di-jet production in DIS as well as in high energy proton-proton 
and proton-nucleus collisions. The present work generalizes the saturation formalism and enlarges 
the transverse momentum range (recall $x$ and $p_t$ are kinematically related) where gluon 
saturation-based models are applied and improve their quantitative accuracy~\cite{cgc-negative}. 
In addition one would also be able to investigate forward-backward (in rapidity) correlations in our 
framework.  

Naturally one expects that our tree-level expression will be renormalized when 
one considers radiative (one-loop) corrections. In analogy with renormalization of 
product of Wilson lines~\cite{cgc-negative,hybrid,djlsv} in small $x$ QCD 
which leads to the JIMWLK/BK evolution equation~\cite{jimwlk,bk}, 
we expect the renormalization of the scattering 
cross section here to lead to a more general evolution equation which incorporates both the 
DGLAP~\cite{dglap} and JIMWLK evolution equations; due to the presence of the large $x$ gluon field 
(not present in saturation formalism) one expects the one loop corrections to result in the DGLAP 
evolution equation in the large $x$ limit. On the other hand and due to presence of the eikonal 
term one would expect to recover the JIMWLK evolution equation in the small $x$ limit. 
Therefore it will be enormously beneficial to calculate the one loop corrections to our result. 
It may also be possible to reformulate this as an effective action approach, analogous to the
McLerran-Venugopalan model~\cite{mv}. If so this would make it possible to treat both 
the early stages in the formation of a Quark Gluon Plasma and the high $p_t$ jet energy loss 
in high energy heavy ion collisions using the same formalism, at least in the earliest times after 
the collision~\cite{kmw}. In summary, the present work takes the first step toward deriving 
a formalism that generalizes the Color Glass Condensate framework by including the physics of 
high $p_t$ and large $x$.

\section*{Acknowledgments}
We acknowledge support by the DOE Office of Nuclear Physics
through Grant No.\ DE-FG02-09ER41620 and by the Idex Paris-Saclay 
though a Jean d'Alembert grant. We would like to thank the staff of 
Pedro Pascual Science Center in Benasque, Spain for their kind hospitality
during the completion of this work. We also thank T. Altinoluk, N. Armesto, 
F. Gelis, E. Iancu, Yu. Kovchegov, A. Kovner, C. Lorc\'e, C. Marquet, 
A.H. Mueller, S. Munier, B. Pire, C. Salgado, G. Soyez, R. Venugopalan, 
D. Wertepny and B. Xiao for critical questions, illuminating discussions 
and helpful suggestions.

\end{document}